\documentclass[10pt,journal]{IEEEtran}

\usepackage{lgrind}
\usepackage{epsfig}
\usepackage{amsmath}    
\usepackage{graphicx}
\usepackage{amssymb}
\pdfoutput=1

\begin{document}

\title{On quantum limit of optical communications: concatenated codes and joint-detection receivers}

\author{Saikat Guha, Zachary Dutton, and Jeffrey H. Shapiro\thanks{Dr. Saikat Guha, member IEEE, and Dr. Zachary Dutton are with the Disruptive Information Processing Technologies Group of
Raytheon BBN Technologies in Cambridge, MA, USA. Prof. Jeffrey H. Shapiro, Fellow IEEE, is with the Research Laboratory of Electronics, Massachusetts Institute of Technology, Cambridge, MA, USA.
(E-mails: sguha@bbn.com; zdutton@bbn.com; jhs@MIT.edu)}}

\maketitle

\begin{abstract}
When classical information is sent over a channel with quantum-state modulation alphabet, such as the free-space optical (FSO) channel, attaining the ultimate (Holevo) limit to channel capacity requires the receiver to make joint measurements over long codeword blocks. In recent work, we showed a receiver for a pure-state channel that can attain the ultimate capacity by applying a single-shot optical (unitary) transformation on the received codeword state followed by simultaneous (but separable) projective measurements on the single-modulation-symbol state spaces. In this paper, we study the ultimate tradeoff between photon efficiency and spectral efficiency for the FSO channel. Based on our general results for the pure-state quantum channel, we show some of the first concrete examples of codes and laboratory-realizable joint-detection optical receivers that can achieve fundamentally higher (superadditive) channel capacity than receivers that physically detect each modulation symbol one at a time, as is done by all conventional (coherent or direct-detection) optical receivers.
\end{abstract}

\maketitle

\section{Introduction}

When a communication channel has a modulation alphabet comprised of quantum states, the Holevo limit is an upper bound to the Shannon capacity of that physical channel paired with any receiver measurement. Even though the Holevo limit is an achievable capacity, the receiver in general must make joint ({\em collective}) measurements over long codeword blocks---measurements that cannot be realized by detecting single modulation symbols followed by classical post processing. This phenomenon of a joint-detection receiver (JDR) being able to yield higher capacity than any single-symbol receiver measurement, is often termed as {\em superadditivity} of capacity\footnote{The more recent usage of the term {\em superadditivity} of capacity refers to a channel with a quantum-state modulation alphabet being able to get a higher classical communications capacity using transmitted states that are entangled over multiple channel uses~\cite{Has09}. For the bosonic channel, we showed that entangled inputs at the transmitter cannot get a higher capacity~\cite{Gio04}. However, one {\em can} get a higher capacity on the bosonic channel by using {\em entangling} (or joint-detection) measurements at the receiver (as opposed to a symbol-by-symbol optical receiver). In this paper, we use the term superadditivity in this latter context. This usage of the term was first adopted by Sasaki, et. al.~\cite{Sas98}.}.

For the pure-loss bosonic channel, laser-light (coherent-state) modulation suffices to attain the Holevo capacity, i.e., use of non-classical transmitted states does not increase capacity~\cite{Gio04}. When Hausladen et. al.'s square-root-measurement (SRM)~\cite{Hau96}, which in general is a positive operator-valued measure (POVM), is applied to a random code it gives the mathematical construct of a receiver measurement that can achieve the Holevo limit. Lloyd et. al.~\cite{Llo10} recently showed a receiver that can attain the Holevo capacity of any quantum channel by making a sequence of ``yes/no" projective measurements on a random codebook. Sasaki et. al.~\cite{Sas98}, in a series of papers, showed several examples of superadditive capacity using pure-state alphabets and the SRM. However, the key practical questions that remain unanswered are how to design modulation formats, channel codes, and most importantly, structured laboratory-realizable designs of Holevo-capacity-approaching JDRs. In~\cite{Guh10a}, we showed that the Holevo limit of a pure-state channel is attained by a projective measurement that can be implemented by a unitary operation (quantum gate) on the codeword state followed by separable projective measurements on the single-modulation-symbol subspaces. 

In this paper, we report on the ultimate  tradeoff between photon efficiency and spectral efficiency of free-space optical communications. We propose a concatenated coding framework for Holevo-capacity-approaching systems, in which the physical joint detection measurement acts on the inner code. Finally, we show concrete examples of codes and JDRs that yield superadditive capacity and high photon efficiency for the optical binary-phase-shift keying (BPSK) signaling alphabet at low photon numbers. These, we believe, are the first structured receiver realizations that exhibit superadditivity, and can be implemented using laboratory optics. 

\section{The multiple-mode lossy optical channel}\label{sec:ultimatecapacity}

Consider a range-$L$ line-of-sight free-space optical (FSO) channel with hard circular transmit and receive apertures of areas $A_t$ and $A_r$ respectively. Assume $\lambda$-center-wavelength quasi-monochromatic transmission. In the near-field propagation regime (Fresnel number product, $D_f \equiv A_tA_r/(\lambda{L})^2 \gg 1$), with no turbulence and no atmospheric extinction, a normal-mode decomposition of the FSO channel yields $M \approx 2D_f$ orthogonal spatio-polarization transmitter-to-receiver modes ($D_f$ spatial modes each of two orthogonal polarizations) with near-unity transmitter-to-receiver power transmissivities ($\eta_m \approx 1$). The orthogonal spatial modes can be thought of as independent parallel channels, in the same sense as conventional RF multiple-input multiple-output (MIMO) channels, where the transmitter and receiver have access to multiple antennas. In the far-field propagation regime ($D_f \ll 1$), only two orthogonal spatial modes (one of each orthogonal polarization) have appreciable power transmissivity ($\eta \approx D_f$ for each mode). Data is modulated using sequences of orthogonal temporal modes (say, flat-top pulses) on each spatial-mode channel. Let us impose---for each spatial mode---a mean transmitted photon number constraint of ${\bar n}_T$ photons per temporal mode (pulse slot), which is also equal to ${\bar n}=\eta_m{\bar n}_T \approx {\bar n}_T$ mean received photon number per temporal mode. The total received mean photon number per pulse slot is $N_R \triangleq M{\bar n}$. For this channel, a coherent-state modulation can attain the ultimate capacity (in bits per pulse slot, or bits/sec/Hz), i.e., the Holevo limit, which is given by~\cite{Gio04}\footnote{Modulation using photon number states can attain the Holevo capacity when the transmissivities $\eta_m =1, \forall m \in \left\{1, \ldots,M\right\}$. When there is loss (i.e., $\eta_m < 1$), such as due to atmospheric extinction, number-state modulation cannot attain the Holevo capacity.}
\begin{equation}
C_{\rm ult}({N_R}) = Mg({\bar n}){\text{ bits/sec/Hz}},
\label{eq:C_ULT_SE}
\end{equation}
where $g(\bar n) \equiv (1+{\bar n})\log_2(1+{\bar n})-{\bar n}\log_2{\bar n}$. Thus, the ultimate limit to the photon information efficiency (PIE), i.e., error-free bits per received photon is given by
\begin{equation}
{C_{\rm ult}(N_R)}/{N_R} ={g({\bar n})}/{\bar n} {\text{ bits/photon}}.
\label{eq:C_ULT_PIE}
\end{equation}
The Holevo limit to capacity can be attained using a coherent-state modulation with transmitted codeword symbols chosen i.i.d. from an isotropic Gaussian prior density with variance ${\bar n}_T$, which translates to a distribution of the received codeword symbols $\left\{|\alpha\rangle\right\}$, $p(\alpha) = {e^{-|\alpha|^2/{\bar n}}}/{\pi{\bar n}}$~\cite{Gio04}. However, achieving the Holevo limit would require a large optimal codebook and a JDR that jointly detects long blocks of codeword symbols. This joint measurement can in principle be performed interchangeably over multiple temporal modes (pulses) or multiple spatial modes (channels). Since the pure-loss $\eta$-transmissivity bosonic channel maps a coherent-state $|\beta\rangle$ to the coherent state $|\sqrt{\eta}\beta\rangle$, we can without loss of generality assume $\eta_m = 1$ for our discussion here. Stated differently, we can always talk in terms of the received states, and impose the mean photon number constraint, $\bar n$ photons per mode, on the received signal.

\begin{figure}
\centering
\includegraphics[width=0.75\columnwidth]{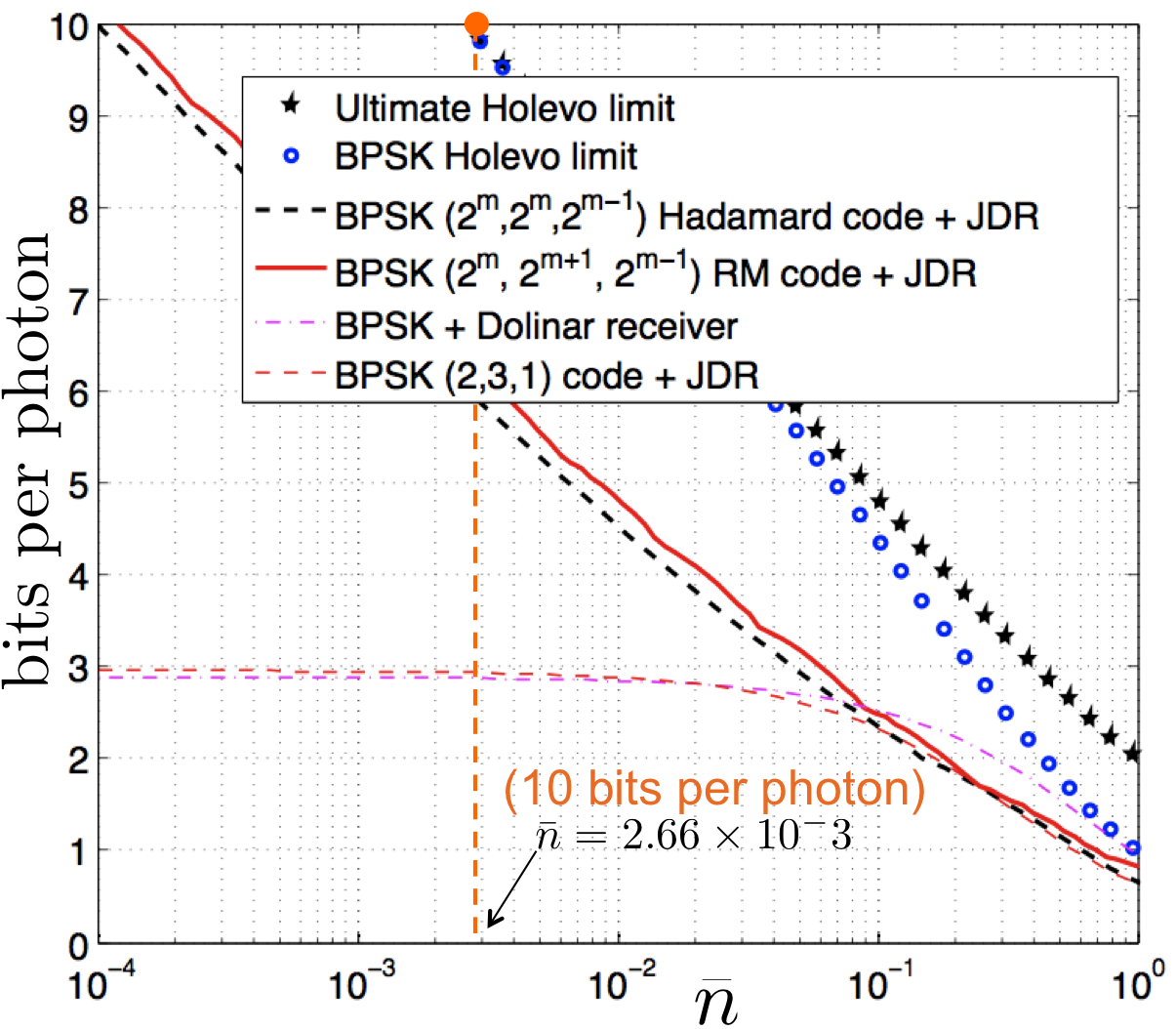}
\vspace{-10pt}
\caption{\small{Photon information efficiency (bits per received photon) as a function of mean photon number per mode, $\bar n$.}}
\label{fig:PIE_vs_nbar}
\end{figure}

\begin{figure}
\centering
\includegraphics[width=0.8\columnwidth]{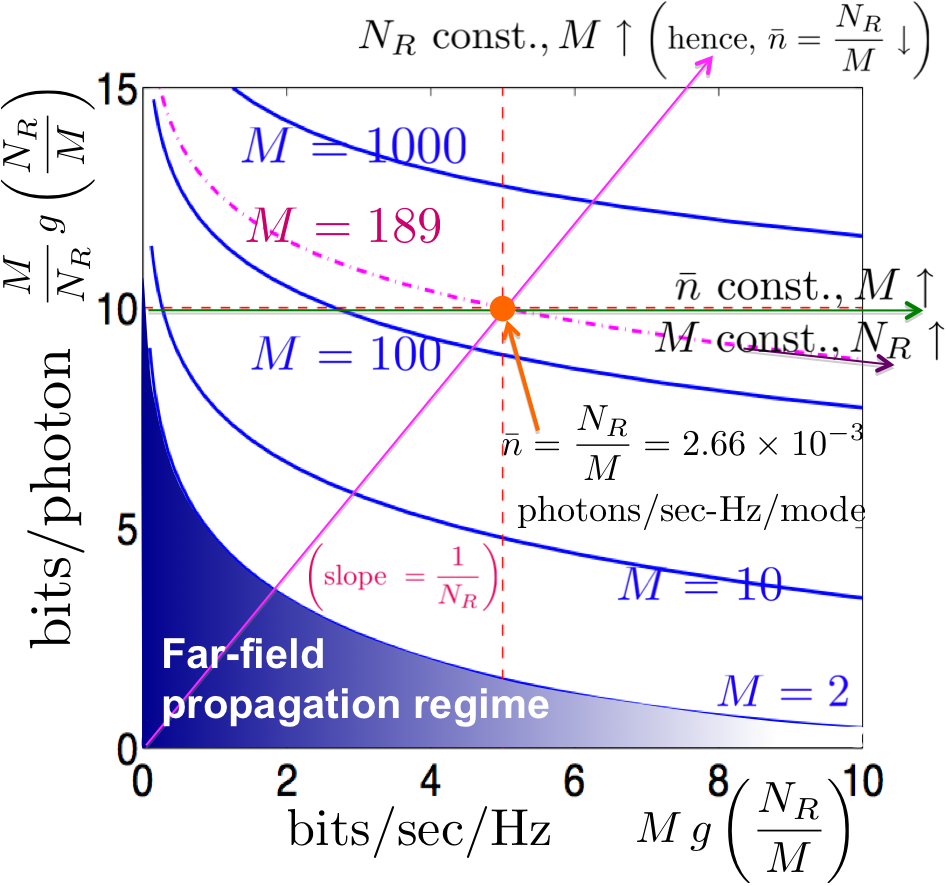}
\caption{Photon information efficiency (PIE) versus spectral efficiency for the multiple-spatial-mode FSO channel.}
\label{fig:PE_vs_SE_ult}
\end{figure}

Figure~\ref{fig:PIE_vs_nbar} plots the ultimate PIE ($g(\bar n)/{\bar n}$) as a function of ${\bar n}$ (with $\bar n$ on a $\log$-scale), and Fig.~\ref{fig:PE_vs_SE_ult} shows the PIE as a function of spectral efficiency for an increasing number ($M$) of spatial modes. In Fig.~\ref{fig:PIE_vs_nbar}, note that with the Holevo-optimal code and receiver, there is no fundamental upper limit to the achievable PIE, but higher PIE necessitates coding with lower mean photon numbers per mode. For instance, a PIE of $10$ bits/photon is achieved at ${\bar n}^*_{\rm ult} = 2.6582 \times 10^{-3}$ photons per mode. From Eqs.~\eqref{eq:C_ULT_SE} and~\eqref{eq:C_ULT_PIE}, we note that the ratio of PIE to spectral efficiency is $1/N_R$. Therefore, in order to attain, say, $10$ bits/photon and $5$ bits/sec/Hz simultaneously, we would need $N_R = 0.5$ mean photon number per pulse slot, and the absolute minimum number of spatial modes needed to meet those photon-efficiency spectral-efficiency numbers, is given by $M_{\rm ult} = \lceil{0.5}/{{\bar n}^*_{\rm ult}}\rceil = 189$. For a $1.55$$\mu$m-wavelength $1$-km-range FSO link, this would imply using ${\sim}7$-cm-radii transmit and receiver apertures  (near-field diffraction-limited operation with $\sim$189 perfect transmit-receiver orthogonal mode pairs) operating at $200$\,MHz modulation bandwidth.   The laws of physics then permit reliable communications at $1$\,Gbps with only $12.8$\,pW of average (and peak) received optical power! 

\section{Attaining the Holevo limit} 

\subsection{Superadditivity in a pure state channel}\label{sec:superadd_purestate}

\begin{figure}
\centering
\includegraphics[width=0.9\columnwidth]{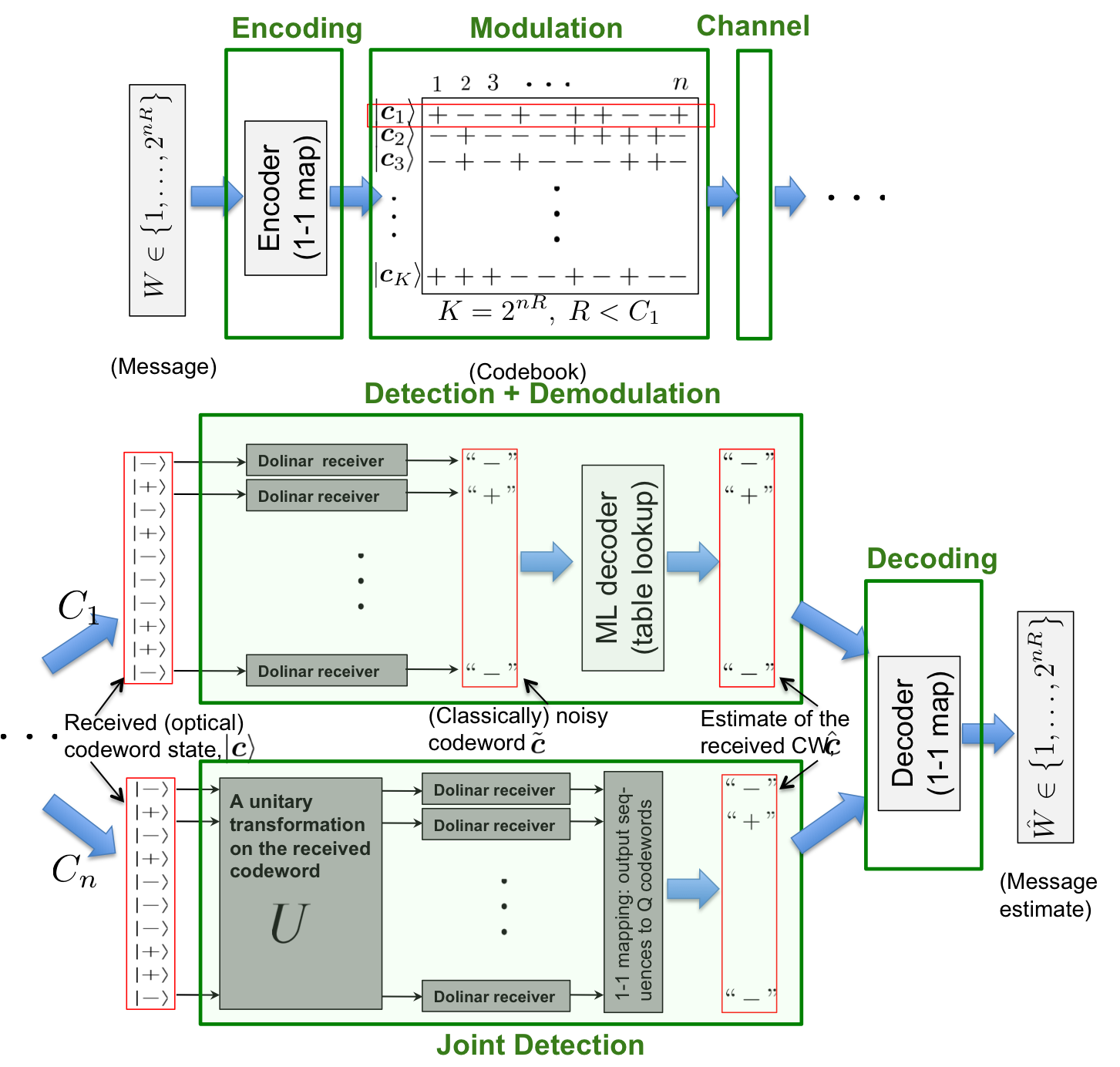}
\vspace{-10pt}
\caption{Classical communication system, shown here for a BPSK alphabet. If the receiver uses symbol-by-symbol detection, maximum capacity $= C_1$ bits/symbol. If the Detection$+$Demodulation block is replaced by a general $n$-symbol joint quantum measurement, maximum capacity $= C_n$ bits/symbol. Superadditivity: $C_\infty > C_n > C_1$, where $C_\infty$ is the Holevo limit. The joint-detection structure shown achieves the Holevo limit for BPSK modulation.}
\label{fig:classicalsystem}
\end{figure}

We encode classical information using a $Q$-ary modulation alphabet of non-orthogonal pure-state {\em symbols} in $S \equiv \left\{|\psi_1\rangle, \ldots, |\psi_Q\rangle\right\}$. Each {\em channel use} constitutes sending one symbol. We assume that the channel preserves the purity of $S$, thus making the transmitted states $\left\{|\psi_q\rangle\right\}$ the ones collected at the receiver. The only source of noise is the physical detection of the states. Assume that the receiver detects each symbol one at a time. Channel capacity is given by the maximum of the single-symbol mutual information, 
\begin{equation}
C_1 = \max_{\left\{p_i\right\}}\max_{\left\{{\hat \Pi}_j^{(1)}\right\}}I_1\left(\left\{p_i\right\}, \left\{{\hat \Pi}_j^{(1)}\right\}\right) {\text{ bits/symbol}},
\label{eq:C1definition}
\end{equation}
where the maximum is taken over priors $\left\{p_i\right\}$ over the alphabet, and over a set of POVM operators $\left\{{\hat \Pi}_j^{(1)}\right\}$, $1 \le j \le J$ on the single-symbol state-space. The measurement of each symbol produces one of $J$ possible outcomes, with conditional probability $P(j|i) = \langle{\psi_i}|{\hat \Pi}_j^{(1)}|\psi_i\rangle$. To achieve reliable communication at a rate close to $C_1$, standard error-correction on the discrete memoryless channel with transition probabilities $P(j|i)$ will be needed. In other words, for any rate $R < C_1$, there exists a sequence of codebooks ${\cal C}_n$ with $K = 2^{nR}$ codewords $|{\boldsymbol c}_k\rangle$, $1 \le k \le K$, each codeword being an $n$-symbol tensor product of states in $S$, and a decoding rule, such that the average probability of decoding error (guessing the wrong codeword), ${\bar P}_e^{(n)} = 1-\frac1K\sum_{k=1}^K{\rm Pr}({\hat k}=k) \to 0,$\footnote{Here, $\hat{k}$ is the receiver's codeword decision, not a Hilbert-space operator.} as $n \to \infty$. In this `Shannon' setting, optimal decoding is a maximum likelihood (ML) decision, which can in principle be pre-computed as a table lookup (see Fig.~\ref{fig:classicalsystem}).
We define $C_n$ as the maximum capacity achievable with measurements that jointly detect up to $n$ symbols. The fact that joint detection may lead to $(n+m)C_{n+m} > nC_n + mC_m$, (or $C_n > C_1$) is referred to as {\em superadditivity} of capacity. The Holevo-Schumacher-Westmoreland (HSW) theorem says that,
\begin{equation}
C_\infty \equiv \max_{\left\{p_i\right\}}S\left(\sum_ip_i|\psi_i\rangle\langle{\psi_i}|\right) = \lim_{n\to \infty}C_n,
\label{eq:HSWpure}
\end{equation}
the Holevo bound, is the ultimate capacity limit, where $S({\hat \rho}) = -{\rm Tr}{\hat \rho}\log_2{\hat \rho}$ is the von Neumann entropy, and that $C_\infty$ is achievable with joint detection over long codeword blocks. Calculating $C_\infty$ however does not require the knowledge of the optimal receiver measurement. In other words, if we replaced the detection and demodulation stages in Fig.~\ref{fig:classicalsystem} by one giant quantum measurement, then for any rate $R < C_\infty$, there exists a sequence of codebooks ${\cal C}_n$ with $K = 2^{nR}$ codewords $|{\boldsymbol c}_k\rangle$, $1 \le k \le K$, and an $n$-input $n$-output POVM over the $n$-symbol state-space $\left\{{\hat \Pi}_k^{(n)}\right\}$, $1 \le k \le K$, such that the average probability of decoding error, ${\bar P}_e^{(n)} = 1-\frac1K\sum_{k=1}^K\langle{\boldsymbol c}_k|{\hat \Pi}_k^{(n)}|{\boldsymbol c}_k\rangle \to 0$, as $n \to \infty$. In~\cite{Guh10a} ({\em Theorem 1}), we showed that for this pure-state channel, a projective measurement can attain $C_\infty$, and can be implemented as a unitary transformation on the codeword's joint quantum state followed by a sequence of separable single-symbol measurements. Finally, note that the fact that joint detection and decoding can help get higher capacity for channels with memory, has been long known. In the $C_n$-achieving receiver shown in Fig.~\ref{fig:classicalsystem}, the classical channel from the $n$-symbol codeword to the detected $n$-symbol block at the output of the separable measurement is not memoryless, even though the physical (quantum) channel is memoryless.

\subsection{Superadditive optical receivers} 

Consider the single-mode pure-loss bosonic channel. The Holevo capacity is given by, 
$
C_{\rm ult}({\bar n}) = g({\bar n})=(1+{\bar n})\log_2(1+{\bar n})-{\bar n}\log_2{\bar n} {\text{ bits/symbol}},
$
where $\bar n$ is the mean photon number per received mode, and it is attained using coherent-state modulation. At high $\bar n$, symbol-by-symbol heterodyne detection asymptotically achieves $C_{\rm ult}({\bar n})$. The low photon number (${\bar n} \ll 1$) regime is the more interesting regime for FSO communications, which is where the joint-detection gain is most pronounced. In Fig.~\ref{fig:PIE_vs_nbar}, we see that binary modulation and coding is sufficient to meet the Holevo limit at low $\bar n$. Specifically, the binary-phase-shift keying (BPSK) alphabet $S_1 \equiv \left\{|\alpha\rangle, |-\alpha\rangle\right\}$, $|\alpha|^2 = {\bar n}$, is the Holevo-optimal binary modulation at ${\bar n} \ll 1$. Dolinar proposed a structured receiver that realizes the binary projective minimum probability of error (MPE) measurement on a pair of coherent states using single-photon detection and coherent optical feedback~\cite{Dol76}. If the Dolinar receiver (DR) is used to detect each symbol, the BPSK channel is reduced to a classical binary symmetric channel (BSC) with capacity $C_1 = 1-H_b(q)$ bits/symbol, where $H_b(\cdot)$ is the binary entropy function and $q = [1-\sqrt{1-e^{-4{\bar n}}}]/2$. This is the maximum achievable capacity when the receiver detects each symbol individually, which includes all conventional (direct-detection and coherent-detection) receivers. The PIE $C_1({\bar n})/{\bar n}$ caps out at $2/\ln 2 \approx 2.89$ bits/photon at ${\bar n} \ll 1$. Closed-form expressions and scaling behavior of $C_n$, the maximum capacity achievable with measurements that jointly detect up to $n$ symbols, for $n \ge 2$ are not known. However, the Holevo limit of BPSK, $C_\infty(\bar n) = H_b([1+e^{-2{\bar n}}]/2)$, can be calculated using Eq.~\eqref{eq:HSWpure}. Good codes and JDRs will be needed to bridge the huge gap between the PIEs $C_1(\bar n)/{\bar n}$ and $C_\infty({\bar n})/{\bar n}$, shown in Fig.~\ref{fig:PIE_vs_nbar}. 

\begin{figure}
\centering
\includegraphics[width=\columnwidth]{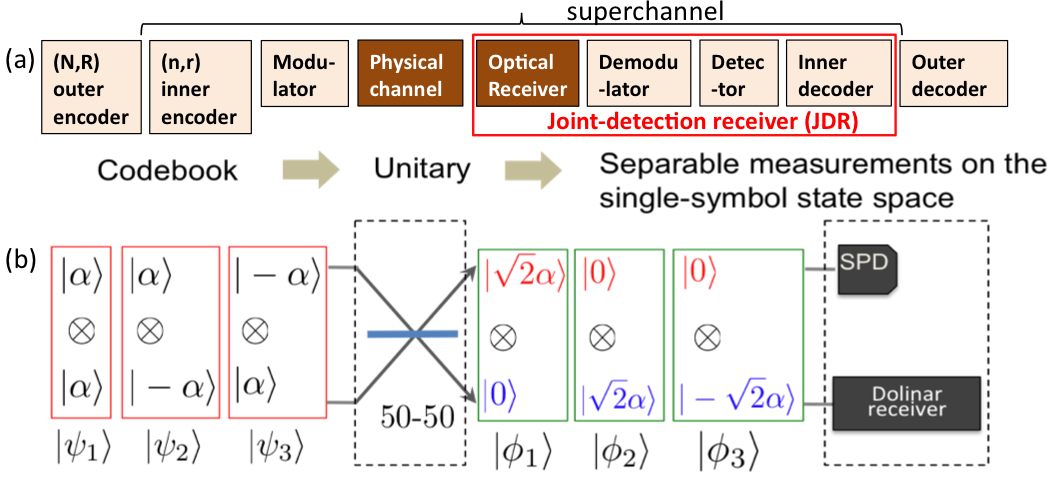}
\vspace{-10pt}
\caption{\small{(a) The ``Forney" concatenated coding architecture~\cite{For66}, in which the {\em channel} is broken up into the physical channel and a receiver. We follow a similar architecture, but with the JDR acting on the inner code. (b) A two-symbol JDR that attains $\approx 2.5\%$ higher capacity for the BPSK alphabet than the best single-symbol (Dolinar) receiver.}}
\label{fig:measurements}
\end{figure}

In principle, as $n \to \infty$, a projective measurement on the codebook that involves an $n$-mode unitary followed by a DR-array is capable of attaining $C_\infty$ without an additional outer code (see Fig.~\ref{fig:classicalsystem})~\cite{Guh10a}. But, in order to construct practical systems with low-complexity reception and decoding, inspired by Forney's early work on capacity-achieving concatenated codes, our code-JDR constructions below will assume an underlying concatenated coding architecture shown in Fig.~\ref{fig:measurements}(a), where we will assume the physical joint-detection part of the receiver acts on the inner codeword. The JDR will in general be allowed to pass on ``errors and erasures" outcome to the outer decoder. In what follows, we will report a few examples of superadditive BPSK code-JDR pairs, where we will quantify superadditivity in terms of the Shannon capacity (or PIE) of the inner superchannel. Detailed calculations with lower and upper error-exponent bounds will be reported elsewhere.

\subsubsection{A two-symbol superadditive JDR}

Some examples of superadditive codes and joint measurements have been reported~\cite{Fuc97, Sas98}, but no structured receiver designs. The simplest non-linear $(2,3,1)$ inner code\footnote{An $(n,K,d)$ binary code has $K$ length-$n$ codewords with minimum Hamming distance $d$. The {\em code rate} is $r = \log_2K/n$.}, containing three of the four $2$-symbol states, $S_2 \equiv \left\{|\alpha\rangle|\alpha\rangle, |\alpha\rangle|-\alpha\rangle, |-\alpha\rangle|\alpha\rangle \right\}$, with priors $(1-2p, p, p)$, $0 \le p \le 0.5$, can attain, with the best $3$-element projective measurement in ${\rm span(S_2)}$, up to $\approx 2.8\%$ higher capacity that $C_1$~\cite{Fuc97}. Since this is a Shannon capacity result, a classical outer code with codewords comprising of sequences of states from $S_2$ will be needed to achieve this capacity $I_2 > C_1$. Using the MPE measurement on $S_2$ (which can be analytically calculated~\cite{Hel76}, unlike the numerically optimized projections in~\cite{Fuc97}), $I_2/C_1 \approx 1.0266$ can be obtained. A receiver that involves mixing the two codeword symbols on a 50-50 beam splitter, followed by a single-photon detector (SPD), and a DR (see Fig.~\ref{fig:measurements}(b)), can attain $I_2/C_1 \approx 1.0249$ (see Fig.~\ref{fig:PIE_vs_nbar}). It is likely that neither of these projective measurements attain $C_2$, because the single-shot measurement that maximizes the accessible information in $S_2$ could in general be a $6$-element POVM~\cite{Sho00}.

\begin{figure}[h]
\centering
\includegraphics[width=\columnwidth]{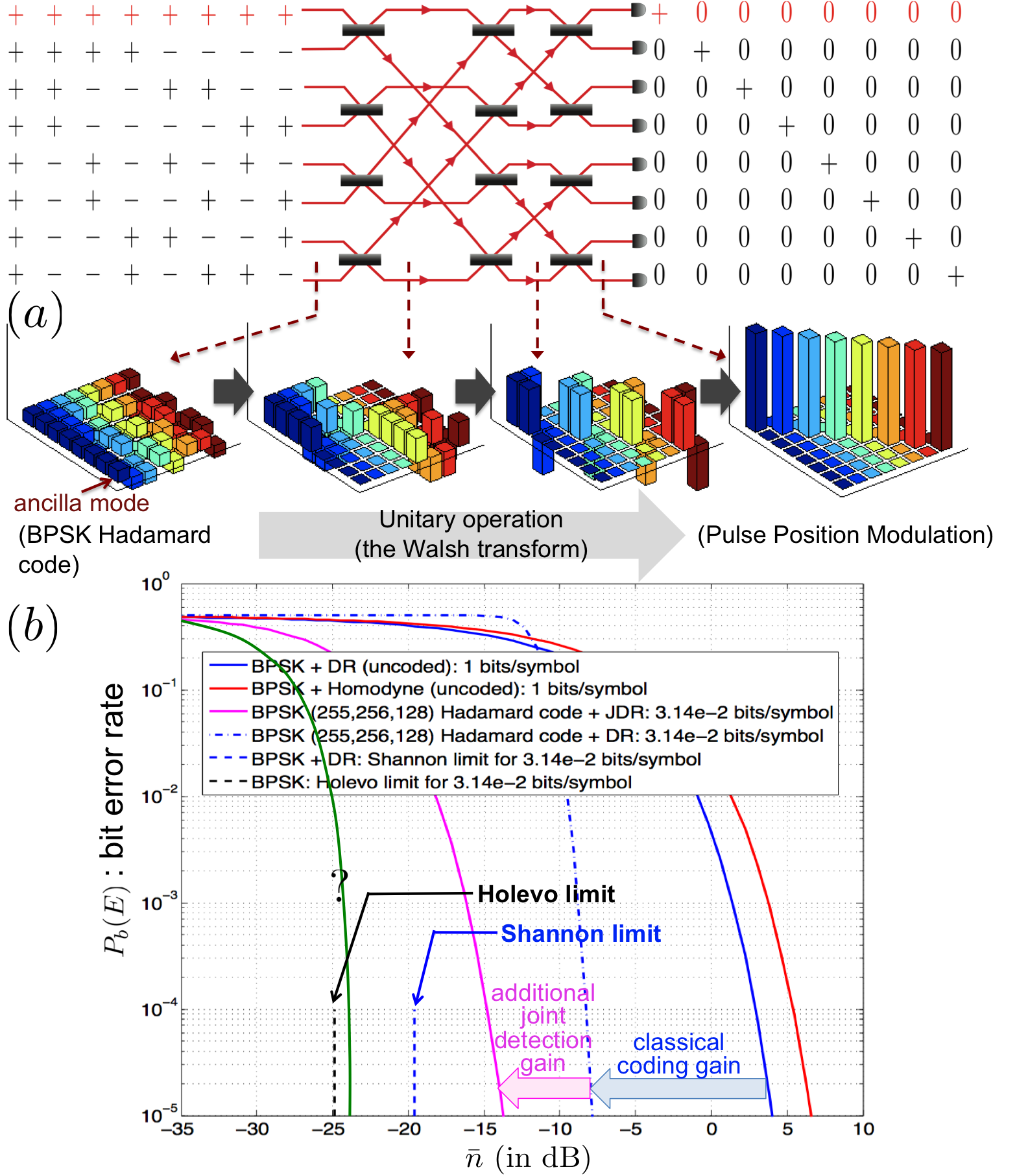}
\vspace{-10pt}
\caption{\small{(a) The BPSK $(7,8,4)$ Hadamard code is unitarily equivalent to the $(8,8,4)$ pulse-position-modulation (PPM) code via a Walsh transform (realized using an all-optical Green Machine) built using $12$ 50-50 beam splitters. (b) Bit error rate plotted as a function of $\bar n$.}}
\label{fig:HadamardJDR}
\end{figure}

\subsubsection{Hadamard code and a superadditive JDR} 

A $(2^m-1,2^m,2^{m-1})$ BPSK Hadamard code, with $\bar n$-mean-photons BPSK symbols is unitarily equivalent to the $(2^m,2^m,2^{m-1})$ pulse-position-modulation (PPM) code---over an underlying on-off-keying (OOK) binary signaling alphabet---with $2^m{\bar n}$-mean-photon-number pulses. The former is slightly more {\em space-efficient}, since it achieves the same equidistant distance profile, but with one less symbol. Consider a BPSK Hadamard code detected by a $2^m$-mode unitary transformation (with one ancilla mode, prepared locally at the receiver, in the $|\alpha\rangle$ state) built using $(n\log_2n)/2$ 50-50 beam splitters arranged in a the ``Green Machine" (GM) format, followed by a separable $n = 2^m$-element SPD-array, as shown (for $n = 8$) in Fig.~\ref{fig:HadamardJDR}(a). The beam splitters `unravel' the BPSK codebook into a PPM codebook, separating the photons into spatially-separate bins. This receiver may be a more natural choice for spatial modulation across $M$ orthogonal spatial modes of a near-field FSO channel. The ancilla mode at the receiver necessitates a local oscillator phase locked to the received pulses, which is hard to implement. Since the number of ancilla modes doesn't scale with the size of the code, we will append the ancilla mode to the transmitted codeword, so that the received ancilla can serve as a pilot tone for our interferometric receiver. The Shannon capacity of this code-JDR superchannel---allowing outer coding over the erasure outcome (no clicks registered at any SPD element)---is $I_n(\bar n) = (\log_2K/K)(1-\exp(-2d{\bar n}))$ bits/symbol. In Fig.~\ref{fig:PIE_vs_nbar}, we plot the envelope, $\max_nI_n(\bar n)/{\bar n}$ (the green dotted plot), as a function of $\bar n$. This JDR not only attains a {\em much} higher superadditive gain than the $n=2$ JDR we described above, it does not need phase tracking and coherent optical feedback like the DR\footnote{Note that $n$-ary PPM signaling also achieves $I_n(\bar n)$ with an SPD, albeit with a much higher ($\times 2^m$) peak power as compared to BPSK. However, {\em Theorem 1} in~\cite{Guh10a} says that the receiver construct shown in Fig.~\ref{fig:classicalsystem} is capable of bridging the rest of the gap to the Holevo limit (i.e., the blue plot in Fig.~\ref{fig:PIE_vs_nbar}) using an optimal BPSK code (minimum peak power) and a JDR.}. In Fig.~\ref{fig:HadamardJDR}(b), we plot the bit error rates $P_b(E)$ as a function of $\bar n$ for uncoded BPSK, and the $(255,256,128)$ BPSK Hadamard code, when detected using a symbol-by-symbol DR and our structured JDR, respectively. The {\em coding gain} now has two components, a (classical) coding gain, and an additional {\em joint-detection gain}. 

\subsubsection{First-order Reed Muller codes and a superadditive JDR} 

\begin{figure}
\centering
\includegraphics[width=0.7\columnwidth]{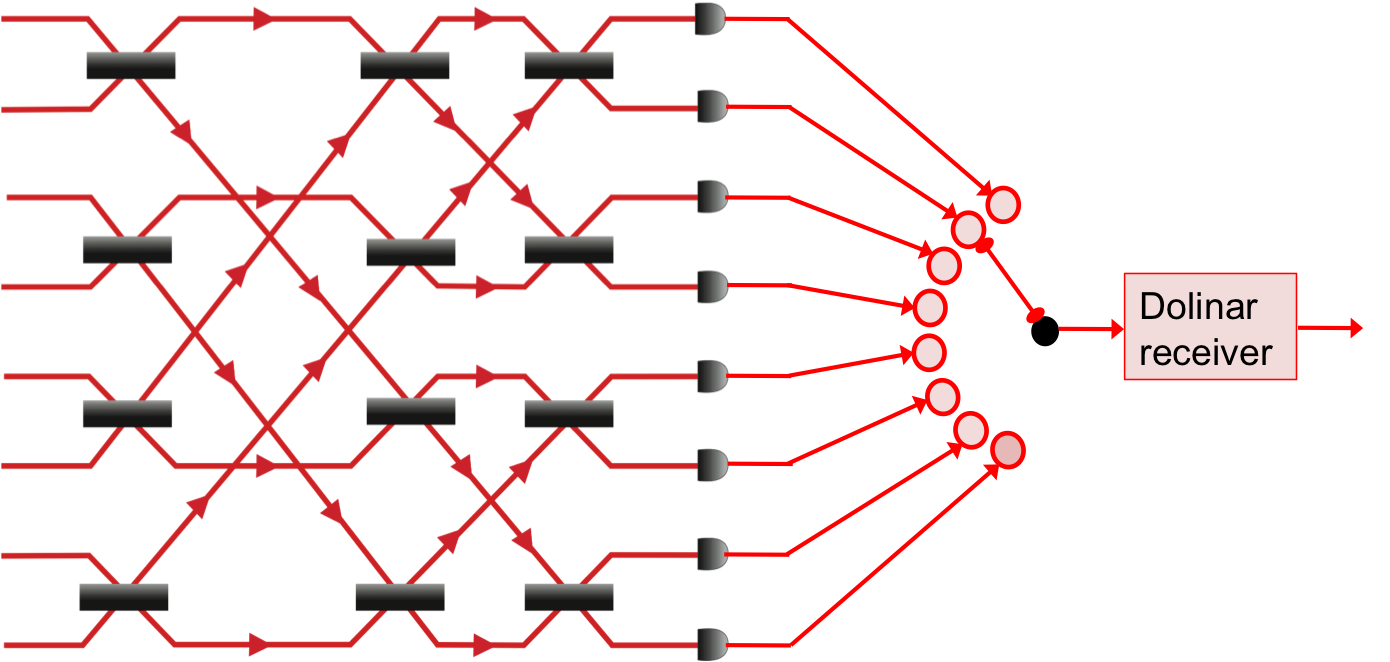}
\caption{Green Machine decoder for the BPSK RM ${\cal R}(1,m)$ code. The first SPD to click sends the remainder of its optical input to the Dolinar receiver.}
\label{fig:reedmullerdecoder}
\end{figure}

Consider the BPSK $(2^m, 2^{m+1}, 2^{m-1})$ first-order Reed Muller (RM) ${\cal R}(1,m)$ code. This is a linear code with the same length as the $(2^m, 2^{m}, 2^{m-1})$ Hadamard code (with the ancilla mode), but twice as many codewords. The ${\cal R}(1,m)$ code can be constructed by appending the $(2^m, 2^{m}, 2^{m-1})$ Hadamard code with each of its $2^m$ codewords with the bits flipped. The Green Machine transforms half of the ${\cal R}(1,m)$ codewords into the $2^m{\bar n}$-mean-photon-number $2^m$-ary PPM codewords, and the other half into the same set of PPM codewords, but with a $\pi$-phase shift. Consider the JDR shown in Fig.~\ref{fig:reedmullerdecoder}, where we detect the output of the GM using an SPD array, to detect which one of the $2^m$ outputs contains the pulse (either $|\sqrt{2^m{\bar n}}\rangle$ or $|-\sqrt{2^m{\bar n}}\rangle$), and the moment one SPD generates a click (confirming presence of a pulse), it switches the remainder of the pulse to a DR to make a decision on its phase. If none of the SPDs click, we call that an erasure outcome (and pass it on to the outer decoder). The Shannon capacity of the above $2^{m+1}$ input $2^{m+1} + 1$ output superchannel is (derivation omitted):
\begin{equation}
I_n(\bar n) = \frac{(1-p_0)(m+1)+H(p_0,1-p_0)-H(p_+, p_-, p_0)}{n},
\end{equation}
where $n=2^m$ is the length of the JDR, $H(\cdot)$ is the Shannon entropy function, $p_0=e^{-{\bar n}_P}$, (${\bar n}_P\equiv 2^m{\bar n}$), and $p_{\pm}=(1-p_0)/2 \pm f({\bar n}_P)$, where $f(\cdot)$ is a definite integral given by
$
f(b) = \frac12{\int_a^1}\sqrt{1-{(a/x)^4}}\,dx, \; a=e^{-b}.
$
The distance symmetry of the ${\cal R}(1,m)$ code can be exploited to compute the Helstrom limit to the MPE measurement on the codewords. The Shannon capacity of the MPE-joint-measurement superchannel is (derivation omitted):
\begin{eqnarray}
I_n(\bar n) &=& \left[(m+1)+a_+^2\log_2a_+^2+a_-^2\log_2a_-^2\right. \nonumber\\
&&\left.+(2^{m+1}-2)c^2\log_2c^2\right]/n,
\end{eqnarray}
where, $c^2 = {(\gamma-\sqrt{\gamma^2-4^mp_0^2})}/{2^{2m+1}}$, $\gamma \equiv 1+2p_0(2^{m-1}-1)+p_0^2$ and $a_{\pm}=\left[{\left(p_0-c^2(2^{m+1}-4)\right)}/{2c} \pm \sqrt{1-p_0^2}\right]/2$.

\begin{figure}
\centering
\includegraphics[width=0.8\columnwidth]{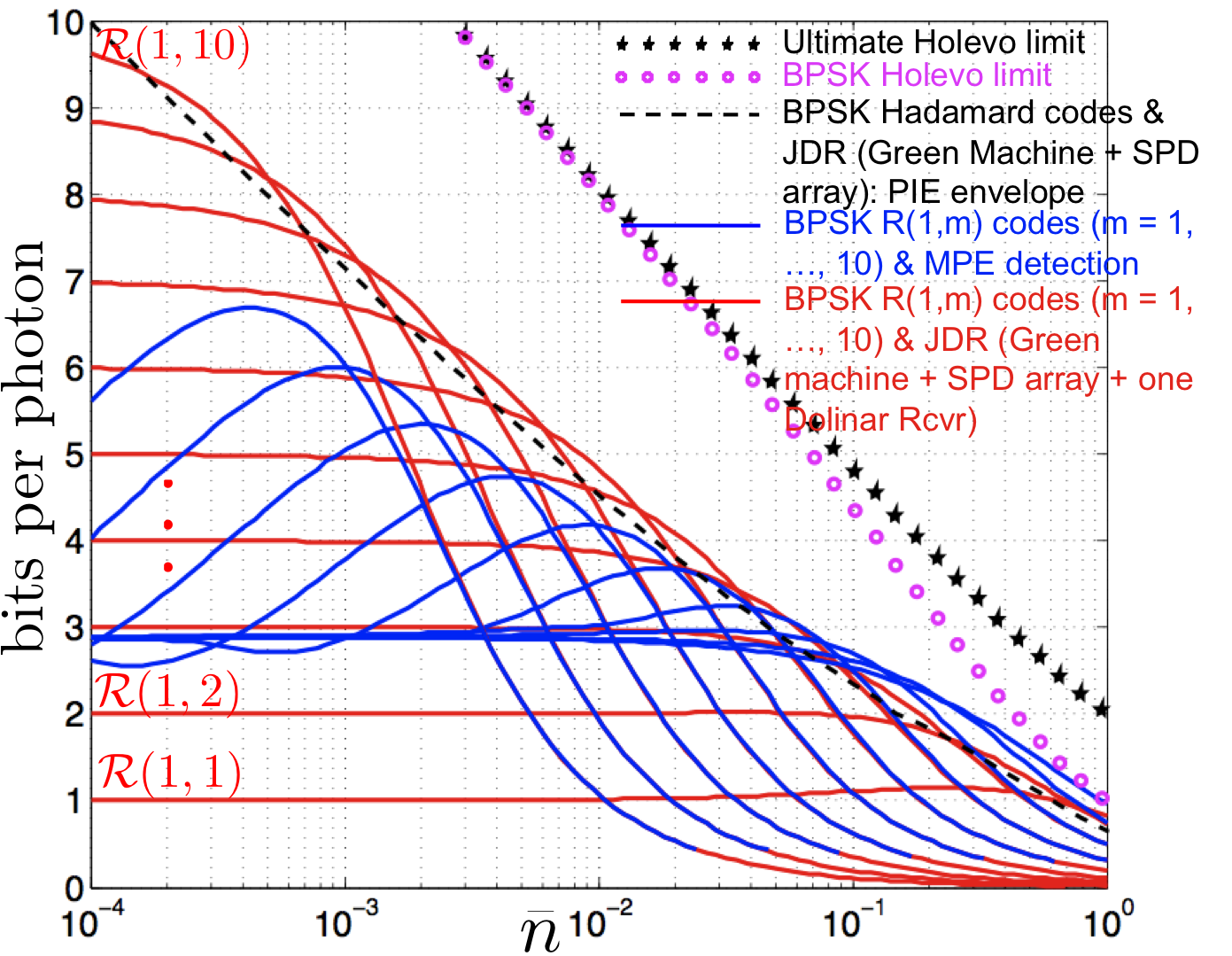}
\caption{PIE as a function of $\bar n$ for BPSK RM ${\cal R}(1,m)$ codes (for $m=1, \ldots, 10$) with the Green Machine JDR shown in Fig.~\ref{fig:reedmullerdecoder} (red plots) and with the Helstrom MPE measurement (blue plots).}
\label{fig:PIE_RMcodes}
\end{figure}

In Fig.~\ref{fig:PIE_RMcodes}, we plot the PIE ($I_n(\bar n)/{\bar n}$) as a function of $\bar n$ for the ${\cal R}(1,m)$ codes ($m=1, \ldots, 10$) with the GM JDR (red plots) and with the MPE measurement (blue plots). It is not surprising that the GM-JDR superchannel can have a higher capacity than that of the equierror channel created by the MPE measurement, which is an ``errors-only" measurement that attains the quantum-minimum average probability of error of discriminating the codewords. The ${\cal R}(1,m)$ family (with the GM JDR) performs slightly better than the Hadamard family (with the GM JDR), and the PIE saturates to $m$ bits per photon at low ${\bar n}$. At low $\bar n$, the PIE attained by the MPE receiver asymptotically approaches that attained by the symbol-by-symbol Dolinar receiver, i.e., $2.89$ bits per photon. A similar result is true also for the Hadamard code. For a finite pure-state-codeword ensemble, the measurement that maximizes the single-shot mutual information is in general different from the MPE measurement (much harder to derive).

\section{Conclusion}
A great deal is known about binary codes that achieve low bit error rates at close to the Shannon limit. It would be interesting to see how close to the Holevo limit can these (or other) codes perform when paired with quantum-limit joint measurements. It will be useful to design codes with symmetries that allow them to approach Holevo capacity, with the unitary $U$ of the JDR in Fig.~\ref{fig:classicalsystem} realizable via a simple network of optical elements along with a low-complexity outer code. The fields of information and coding theory have had a rich history. Even though many ultimate limits were determined in Shannon's founding paper, it took generations of magnificent coding theory research, to ultimately find practical capacity-approaching codes. Likewise, realizing reliable communications on an optical channel close to the Holevo limit at high photon and spectral efficiencies might take a while, it certainly does seem to be visible on the horizon.

\vspace{5pt}
This work was supported by the DARPA Information in a Photon (InPho) program, contract\# HR0011-10-C-0159. Discussions with Prof. Lizhong Zheng, MIT and Dr. Mark Neifeld, DARPA are gratefully acknowledged.

\end{document}